\documentclass[prl,aps,twocolumn,floatfix,superscriptaddress,letterpaper,longbibliography]{revtex4-2}
\usepackage{graphicx}
\usepackage{bm}
\usepackage{amsmath,amsfonts}
\usepackage{amsbsy}
\usepackage[colorlinks=true,linkcolor=blue,urlcolor=blue,citecolor=blue,breaklinks=true]{hyperref}
\usepackage[utf8]{inputenc}
\DeclareUnicodeCharacter{03BC}{\mbox{$\mu$}}
\DeclareUnicodeCharacter{2009}{ }

\begin{document}

\title{Viscous Properties of a Degenerate One-Dimensional Fermi Gas}
\author{Wade DeGottardi}

\affiliation{Institute for the Research in Electronics and Applied Physics, University of Maryland, College
Park, Maryland 20742, USA}

\affiliation{Joint Quantum Institute, NIST/University of Maryland, College Park, Maryland, 20742, USA}

\author{K. A. Matveev}

\affiliation{Materials Science Division, Argonne National Laboratory,
  Argonne, Illinois 60439, USA}

\date{\today}

\begin{abstract}
  We study the viscous properties of a system of
  weakly interacting spin-$\frac{1}{2}$ fermions in one dimension. Accounting for the
  effect of interactions on the quasiparticle energy spectrum, we obtain the bulk
  viscosity of this system at low temperatures. Our result is valid for frequencies that
  are small compared with the rate of fermion backscattering. For
  frequencies larger than this exponentially small rate, the
  excitations of the system become decoupled from the center of mass
  motion, and the fluid is described by two-fluid hydrodynamics. We
  calculate the three transport coefficients required to describe
  viscous dissipation in this regime.
\end{abstract}

\maketitle

Hydrodynamics is a classical description of the mechanical and thermal properties of a fluid near
equilibrium~\cite{landau_fluid_1987}. The application of hydrodynamics to low-dimensional quantum liquids has
provoked a great deal of excitement and has important implications for both
experiment~\cite{joseph_observation_2011,bandurin_negative_2016,crossno_observation_2016,moll_evidence_2016} and
theory~\cite{levitov_electron_2016,lucas_transport_2016,guo_higher-than-ballistic_2017}. Applied to the
collective behavior of electrons in quantum wires or carbon nanotubes, hydrodynamics would potentially offer new
insight into the transport properties of these systems~\cite{andreev_hydrodynamic_2011,matveev_two-fluid_2019}.
From a theoretical perspective, reconciling the behavior of a one-dimensional (1D) quantum liquid with that of a
classical dissipative fluid raises important questions.

In particular, many quantities of interest are directly related to dissipation. The dissipative dynamics of a 1D
fluid is characterized by two transport coefficients, the thermal conductivity and the bulk viscosity.
Ultimately, the calculation of these parameters requires input from a microscopic model. Much progress has been
made in understanding 1D systems through the study of integrable models~\cite{korepin_quantum_1997}. However,
integrability precludes the relaxation of excitations and thus these models are incapable of accounting for
dissipative effects in real fluids~\cite{sutherland_beautiful_2004,imambekov_one-dimensional_2012}. For the
particular case of spinless quantum liquids in one dimension, however, the bulk viscosity and thermal
conductivity can be evaluated~\cite{matveev_viscous_2017,degottardi_electrical_2015} within the
Tomonaga-Luttinger liquid framework~\cite{haldane_`luttinger_1981,giamarchi_quantum_2004}.

One-dimensional systems of spin-$\frac{1}{2}$ fermions generally relax much more rapidly than their spinless
counterparts. Thus, the results~\cite{matveev_viscous_2017,degottardi_electrical_2015} for the transport
coefficients of spinless systems do not apply to those with spin. Given the importance of systems of
spin-$\frac{1}{2}$ fermions,
a means of calculating their transport coefficients would be beneficial. Unfortunately, spin-charge
separation~\cite{dzyaloshinskii_correlation_1974,schmidt_spin-charge_2010,giamarchi_quantum_2004} frustrates the
application of the Tomonaga-Luttinger liquid theory for this purpose. On the other hand, the regime in which the
transport coefficients are largest is in fact theoretically accessible. Indeed, transport coefficients are
proportional to the relaxation time of the system~\cite{pitaevskii_physical_1981} and are thus large for weakly
interacting systems. Furthermore, as long as the relevant energy scale---in this case temperature---is large
compared with the interactions, the effect of spin-charge separation can be neglected~\cite{karzig_energy_2010}.
For these reasons, in this Letter we study the case of weakly interacting spin-$\frac{1}{2}$ fermions.

The transport coefficients of 1D systems reflect their unique relaxation properties.
A 1D Fermi gas at low temperatures exhibits two disparate relaxation
rates~\cite{micklitz_transport_2010}. The slowest relaxation process
involves backscattering of particles, in which, say, a right mover is
converted to a left mover. For such a process to occur, a hole must
pass through the bottom of the band. Hence, the rate for these
processes takes the activated form $1/\tau \sim e^{-E_F/T}$, where
$E_F$ is the Fermi energy and $T$ is the temperature. On the other
hand, typical particle-hole excitations relax much more rapidly, with
a characteristic rate $1/\tau_{\rm{ex}}$ that scales as a power of
$T$.

Transport coefficients are associated with specific perturbations of
the system. The thermal conductivity $\kappa$ quantifies the heat
current that arises from the application of a temperature
gradient. Since currents in one dimension are associated with an
imbalance between the right and left movers, the thermal conductivity
is dominated by fermion backscattering, i.e.,
$\kappa \propto
\tau$~\cite{micklitz_transport_2010,degottardi_electrical_2015,matveev_thermal_2019,samanta_thermal_2019}. The
bulk viscosity $\zeta$, on the other hand, is a measure of the amount
of entropy generated by a change in the fluid density---a perturbation
that affects right and left movers equally. This perturbation creates
particle-hole excitations and thus
$\zeta \propto
\tau_{\rm{ex}}$~\cite{matveev_viscous_2017,degottardi_equilibration_2019}. The
calculation of $\zeta$ is, in general, more challenging than that of
$\kappa$ since particle-hole excitations exhibit a spectrum of
relaxation rates whereas $\kappa$ is dominated by a single
rate~\cite{matveev_thermal_2019}. The study of the bulk viscosity of a 1D gas of spin-$\frac{1}{2}$
particles is our main goal. Taken together with an
understanding of thermal transport, our results give a
complete hydrodynamic description of a 1D gas of spin-$\frac{1}{2}$
fermions.

The vast majority of experimentally relevant 1D systems of fermions, including quantum wires in GaAs as well as cold atomic gases, exhibits single particle energy spectra that are quadratic in momentum. It is well known that a straightforward calculation of the bulk viscosity in this case gives $\zeta = 0$~\cite{pitaevskii_physical_1981,sykes_transport_1970}. This presents a theoretical challenge which we overcome by properly accounting for the renormalization of the energy spectrum by the interactions.

Classical hydrodynamics describes the properties of the system at
frequencies $\omega$ that are small compared with the slowest
relaxation rate. In our system, the slowest relaxation process is
fermion backscattering and therefore this condition is
$\omega \ll 1/\tau$. The presence of two disparate scales of relaxation rates,
$1/\tau$ and $1/\tau_{\rm{ex}}$, ensures that there exists a broad
frequency range $1/\tau \ll \omega \ll 1/\tau_{\rm{ex}}$. In this
regime, particle hole excitations, due to their rapid equilibration,
behave as a gas moving with a well-defined velocity. In the absence of
backscattering, this velocity can differ from the velocity of the
center of mass. Therefore, the system is properly described by
two-fluid hydrodynamics, similar to the theory developed for
superfluid $^4$He~\cite{khalatnikov_introduction_2000}. This
conclusion is central to recent theoretical work on the
superfluid-like behavior displayed by 1D liquids at finite
frequencies~\cite{matveev_second_2017,matveev_propagation_2018}. In
two-fluid hydrodynamics, bulk viscosity is described by three
transport coefficients. We obtain analytic expressions for these
quantities.

We start by considering the thermodynamic equilibrium state of a
noninteracting 1D Fermi gas.  In the absence of a magnetic field the
spins are degenerate, and the occupation numbers of the fermion states
depend only on the momentum,
\begin{equation}
  \label{eq:Fermi_distribution}
  n_p^{(0)}=\frac{1}{\exp\left(\frac{\epsilon_p-up-\mu}{T}\right)+1}.
\end{equation}
Here $\epsilon_p=p^2/2m$ is the energy of the fermion with momentum
$p$, while $\mu$ is the chemical potential.  The appearance of the
term $-up$ in Eq.~(\ref{eq:Fermi_distribution}) is dictated by the
conservation of momentum in a uniform system. The physical meaning of the parameter $u$ is the velocity of the
gas.

In the presence of an infinitesimal gradient of velocity
$\partial_x u$, weak interactions in the Fermi gas lead to scattering
of particles, resulting in dissipation.  The power $W$ dissipated in
the system is
\begin{equation}
  \label{eq:power_dissipation}
  W=\zeta L(\partial_x u)^2,
\end{equation}
where $\zeta$ is the bulk viscosity and $L$ is the system size~\cite{landau_fluid_1987}. Below
we use Eq.~(\ref{eq:power_dissipation}) to evaluate $\zeta$.

To obtain the power dissipated in the Fermi gas we employ the standard
expression for the entropy $S=-2\sum_p[n_p\ln n_p+(1-n_p)\ln (1-n_p)]$
in terms of the occupation numbers $n_p$.  Differentiation of $S$ with
respect to time gives the entropy production rate
\begin{equation}
  \label{eq:entropy_production}
  \dot S=-2\sum_p\dot  n_p\ln\frac{n_p}{1-n_p}.
\end{equation}
Substitution of the unperturbed occupation numbers
(\ref{eq:Fermi_distribution}) for $n_p$ within the logarithm in
Eq.~(\ref{eq:entropy_production}) results in $\dot S=0$ by virtue of the
laws of conservation of the number of particles, momentum, and
energy.  On the other hand, a small velocity gradient generates a
correction to the occupation numbers $\delta n_p\propto\partial_x u$.
Substituting $n_p=n_p^{(0)}+\delta n_p$ into
Eq.~(\ref{eq:entropy_production}) and expanding to first order in
$\delta n_p$, one finds the dissipation rate $W=T\dot S$ in the form
\begin{equation}
  \label{eq:W}
  W=-2T\sum_p\frac{\dot n_p\delta n_p}{n_p^{(0)}\big(1-n_p^{(0)}\big)}.
\end{equation}
Both $\dot n_p$ and $\delta n_p$ are proportional to the perturbation
$\partial_x u$.  Therefore $W\propto (\partial_x u)^2$, as expected
from Eq.~(\ref{eq:power_dissipation}).

We will obtain $\dot n_p$  and $\delta n_p$ by using the Boltzmann
equation, which can be written as a combination of the relation
\begin{equation}
  \label{eq:Boltzmann}
  \dot n_p=\partial_tn_p + (\partial_p\epsilon_p)\partial_x n_p
\end{equation}
with the expression for $\dot n_p$ in terms of the collision integral,
$\dot n_p=I[n_p]$.  Because the correction $\delta n_p$ is small, to
leading order one can use unperturbed occupation numbers $n_p^{(0)}$
in the right-hand side of Eq.~(\ref{eq:Boltzmann}). The correction
$\delta n_p$ should then be found by solving
$\dot n_p = I[n_p^{(0)} + \delta n_p]$.

A nonzero gradient of velocity $\partial_xu$ of the gas results in a
time-dependent density of particles, which in turn gives rise to a
time-dependent chemical potential $\mu(t)$ and temperature $T(t)$.
Keeping this in mind, substitution of the unperturbed occupation
numbers (\ref{eq:Fermi_distribution}) into Eq.~(\ref{eq:Boltzmann})
yields
\begin{equation}
  \label{eq:dot-n_unperturbed}
  \dot n_p =\frac{1}{T} n_p^{(0)}\Big(1-n_p^{(0)}\Big)\!
  \bigg[
    \frac{\partial_tT}{T}
    \bigg(
      \frac{p^2}{2m}-\mu
    \bigg)\!
    +\partial_t\mu
    +\frac{p^2}{m}\partial_xu
  \bigg].
\end{equation}
Here for simplicity we have chosen a point in space where $u=0$.
Because the collisions of particles leading to the equilibration of
the system conserve particle number and energy, $\dot n_p$ must
satisfy the conditions
\begin{equation}
  \label{eq:conditions}
  \int \dot n_p dp=0,
  \quad
  \int \epsilon_p\dot n_p dp=0.
\end{equation}
These constraints enable one to obtain the rates of change of the
temperature and chemical potential,
\begin{equation}
  \label{eq:time-derivatives}
  \frac{\partial_tT}{T}=\frac{\partial_t\mu}{\mu}=-2\partial_x u.
\end{equation}
The substitution of Eq.~(\ref{eq:time-derivatives}) into
Eq.~(\ref{eq:dot-n_unperturbed}) yields $\dot n_p=0$.

This conclusion implies that the dissipation rate (\ref{eq:W}) vanishes,
and thus the bulk viscosity $\zeta=0$.  This can be understood as
follows \cite{matveev_viscous_2017}.  Consider a system with a fixed
number of particles $N$ in a box of size $L(t)$ such that $N/L=n$.
From the standard continuity equation for the particle density $n$ we
conclude that its time dependence is given by
$(\partial_t n)/n=-\partial_x u$.  Then the gradient of velocity can
be related to the time derivative of the system size,
$(\partial_t L)/L=\partial_x u$.  Given that the energy levels
$\epsilon_p=p^2/2m$ are multiples of $(2\pi\hbar/L)^2/2m$, we conclude
that $(\partial_t \epsilon_p)/\epsilon_p=-2\partial_x u$.  Equation
(\ref{eq:time-derivatives}) then indicates that the temperature and
chemical potentials change at the same rate as $\epsilon_p$, and the
ratio $(\epsilon_p-\mu)/T$ in the Fermi-Dirac distribution
Eq.~(\ref{eq:Fermi_distribution}) for $u=0$ remains unchanged.  Thus,
the perturbation $\partial_x u$ does not drive the system out of
equilibrium, resulting in no dissipation and $\zeta=0$.  An analogous
result for a classical ideal gas is well known
\cite{pitaevskii_physical_1981}.

The above argument requires that the particle energies scale as
$\epsilon_p\propto p^2$ (or, more precisely, as any power of $|p|$).
In an interacting system the fermion energies are affected by other
particles and the scaling $\epsilon_p\propto p^2$ no longer holds. To account for this effect, we
consider the usual two-particle interactions described by the
Hamiltonian
\begin{equation}
  \label{eq:interaction_hamiltonian}
  \hat V = \frac{1}{2L}\sum_{\substack{p,p',q\\ \sigma,\sigma'}}
  V(q)a_{p+q,\sigma}^\dagger a_{p'-q,\sigma'}^\dagger
      a_{p',\sigma'}^{}a_{p,\sigma}^{}.
\end{equation}
Here $V(q)$ is the Fourier transform of the interaction potential and
$a_{p,\sigma}^{}$ is the annihilation operator of a fermion with
momentum $p$ and $z$-component of spin $\sigma$.  Assuming that the
interactions are weak, we will limit our treatment to first order
perturbation theory in $\hat V$.  In this approximation, the energy of
the state with occupation numbers $n_{p,\sigma}$ has the form
\begin{equation}
  \label{eq:E}
  E=\sum_{p,\sigma}\frac{p^2}{2m}n_{p,\sigma}
  +\frac{1}{2L}\sum_{\substack{p,p'\\ \sigma,\sigma'}}
  [V(0)-V(p-p')\delta_{\sigma,\sigma'}]n_{p,\sigma}n_{p',\sigma'}.
\end{equation}
Since the energy of the many-body state is a functional of the
occupation numbers $n_{p,\sigma}$, the quasiparticle energies can be
obtained as $\epsilon_{p,\sigma} = \delta E/\delta n_{p,\sigma}$, resulting in
\begin{equation}
  \label{eq:quasiparticle_energies}
  \epsilon_p=\frac{p^2}{2m}+
  \int \frac{dp'}{2\pi\hbar}[2V(0)-V(p-p')]n_{p'},
\end{equation}
where we have assumed spin degeneracy and omitted the spin indices.
The energy spectrum (\ref{eq:quasiparticle_energies}) is no longer
quadratic in $p$.  For a generic interaction, this will result in
non-vanishing $\dot n_p$, which we will evaluate to first order in
$\hat V$.

It is worth noting that the low-energy properties of one-dimensional systems of interacting
fermions are usually described within the framework of Luttinger
liquid theory \cite{giamarchi_quantum_2004}, in which the elementary
excitations have bosonic statistics.  On the other hand, it was shown
in Ref.~\cite{karzig_energy_2010} that for weak interactions the
curvature of the spectrum suppresses the Luttinger liquid effects, and
the simple perturbative treatment of interactions is appropriate.  For
particles with energies of the order of $T$, the criterion of
Ref.~\cite{karzig_energy_2010} is $p_FV(0)/\hbar\ll T$, where
$p_F=\sqrt{2m\mu}$ is the Fermi momentum
\cite{matveev_relaxation_2020}. To account for the effects of
interactions in the Boltzmann equation formalism, we notice that the
first-order expressions for the energy of the system (\ref{eq:E}) and
quasiparticle energy (\ref{eq:quasiparticle_energies}) are consistent
with Fermi liquid theory \cite{lifshitz_statistical_1980}.  The
evaluation of the transport coefficients in this approach was
performed in Refs.~\cite{abrikosov_theory_1959, sykes_transport_1970}.
Below, we simplify and adapt the evaluation
\cite{sykes_transport_1970} of $\dot n_p$ induced by a small gradient
of velocity to the case of one dimension and weak interactions.

To proceed, we observe that
Eqs.~(\ref{eq:Fermi_distribution})--(\ref{eq:Boltzmann}) are still
applicable, provided that the quasiparticle energies $\epsilon_p$
include the Fermi liquid corrections \cite{abrikosov_theory_1959,
  sykes_transport_1970}.  Evaluation of $\dot n_p$ should now allow
for the possibility of $\epsilon_p$ depending on $T$ and $\mu$, which
enter via the occupation numbers in
Eq.~(\ref{eq:quasiparticle_energies}).  Then, substitution of
Eq.~(\ref{eq:Fermi_distribution}) for $n_p$ in the right-hand side of
Eq.~(\ref{eq:Boltzmann}) yields
\begin{eqnarray}
  \label{eq:dot-n}
  \dot n_p& =&\frac{1}{T} n_p^{(0)}\Big(1-n_p^{(0)}\Big)\!
    \bigg[
    \bigg(
    \frac{\epsilon_p-\mu}{T}
    -\frac{\partial\epsilon_p}{\partial T}
    \bigg)
    \partial_tT
\nonumber\\
          &&
    +
    \bigg(
    1-\frac{\partial\epsilon_p}{\partial \mu}
             \bigg)
             \partial_t\mu
    +p(\partial_p\epsilon_p)\partial_xu
  \bigg].
\end{eqnarray}
We now substitute Eq.~(\ref{eq:quasiparticle_energies}) for
$\epsilon_p$ and obtain $\dot n_p$ in linear order in the interaction
potential.  The values of time derivatives $\partial_t T$ and
$\partial_t \mu$ are fixed by the conservation laws
(\ref{eq:conditions}).  For quasiparticles with energies near the
Fermi level, $|\epsilon_p-\mu|\sim T$, to leading order in temperature we
find
\begin{equation}
  \label{eq:dot-n_result}
  \dot n_p=\frac{\gamma}{4\mu T} n_p^{(0)}\Big(1-n_p^{(0)}\Big)\!
               \bigg[
               v_F^2(|p|-p_F)^2-\frac{\pi^2 T^2}{3}
               \bigg]
               \partial_xu.
\end{equation}
Here the dimensionless parameter
\begin{equation}
  \label{eq:eta}
  \gamma=\frac{V(0)-V(2p_F)+2p_FV'(2p_F)-2p_F^2V''(2p_F)}{2\pi\hbar v_F}
\end{equation}
characterizes the strength of interactions and $v_F=\sqrt{2\mu/m}$ is
the Fermi velocity.

In order to obtain the dissipation rate (\ref{eq:W}), one should find
a small correction $\delta n_p$ to the equilibrium distribution
function (\ref{eq:Fermi_distribution}) by inverting the collision
integral: $\dot n_p = I[n_p^{(0)} + \delta n_p]$.  For small
$\delta n_p\propto \partial_x u$, the latter can be linearized.  The
linearized collision integral for 1D spin-$\frac12$ fermions was
studied in Ref.~\cite{matveev_relaxation_2020}.  Remarkably, in the
low-temperature regime the correction to $n_p^{(0)}$ with momentum
dependence of Eq.~(\ref{eq:dot-n_result}) is an eigenmode of the
collision integral, with the relaxation rate
\begin{equation}
  \label{eq:rate}
  \frac{1}{\tau_2}=\frac{9[V(0)V(2p_F)-V(2p_F)^2-2p_F V(0)V'(2p_F)]^2}
  {64\pi^3\hbar^5 v_F^4}T.
\end{equation}
The latter statement means that to leading order in $T/\mu\ll 1$ the
naive relaxation time approximation $\dot n_p=-\delta n_p/\tau_2$ is
exact.

Next, we substitute $\delta n_p=-\tau_2 \dot n_p$ and
Eq.~(\ref{eq:dot-n_result}) into the expression (\ref{eq:W}) for the
dissipation rate and use Eq.~(\ref{eq:power_dissipation}) to obtain
the bulk viscosity
\begin{equation}
  \label{eq:zeta_result}
  \zeta=\frac{2\pi^3}{45}\,
  \frac{\gamma^2T^4\tau_2}{\hbar v_F\mu^2}.
\end{equation}
This result in combination with Eqs.~(\ref{eq:eta}) and
(\ref{eq:rate}) gives a microscopic expression for the bulk viscosity
of the degenerate 1D gas of spin-$\frac12$ fermions.  Given the
temperature dependence of the relaxation time $\tau_2\propto 1/T$, we
conclude that $\zeta\propto T^3$.

Our result (\ref{eq:zeta_result}), derived assuming a time-independent
perturbation $\partial_x u$, is applicable at frequencies
$\omega\ll 1/\tau$.  We now consider the bulk viscosity of the system
at frequencies in the range $1/\tau\ll\omega\ll1/\tau_{\rm ex}$, where
the backscattering rate is exponentially small,
$1/\tau\propto e^{-E_F/T}$, and the quasiparticle relaxation rate
$1/\tau_{\rm ex}=1/\tau_2\propto T$.  As discussed above, in this
regime the system is described by two-fluid hydrodynamics originally
developed for superfluid $^4$He \cite{khalatnikov_introduction_2000}
and adapted to one dimension \cite{matveev_propagation_2018}.  The
rate of viscous dissipation in this theory is controlled by three
transport coefficients, $\zeta_1$, $\zeta_2$, and $\zeta_3$,
\begin{equation}
  \label{eq:W_two-fluid}
  \frac{W}{L}=\zeta_2(\partial_x v_n)^2
  +\zeta_3[\partial_x(j-\rho v_n)]^2
  +2\zeta_1[\partial_x(j-\rho v_n)](\partial_x v_n).
\end{equation}
Here, $v_n$ is the velocity of the normal component of the fluid, $j$
is the mass current, and $\rho$ is the mass density.

To obtain microscopic expressions for the bulk viscosities in
Eq.~(\ref{eq:W_two-fluid}) for the 1D Fermi gas, we first notice that
in the two-fluid regime one can assume $1/\tau=0$, thereby neglecting
the backscattering of fermions.  Then, the numbers of the right- and
left-moving fermions are conserved, and instead of $\mu$, the
occupation numbers are described by two chemical potentials
$\mu_{R,L}=\mu\pm\delta\mu/2$,
\begin{equation}
  \label{eq:Fermi_distribution_two-fluid}
  n_p^{(0)}=\frac{1}{\exp\left(\frac{\epsilon_p-up-\mu-(\delta\mu/2){\rm
        sgn\,}p}{T}\right)+1}.
\end{equation}
For $\delta\mu\neq0$, the center of mass velocity of the Fermi gas is
different from the velocity $u$ of the gas of elementary excitations.

Next, we relate the parameters of the distribution function
(\ref{eq:Fermi_distribution_two-fluid}) to $v_n$ and $j$ in
Eq.~(\ref{eq:W_two-fluid}).  The gas of particle-hole excitations
plays the role of the normal component of the fluid
\cite{matveev_second_2017, matveev_propagation_2018}, and thus
$v_n=u$.  Then, using Eq.~(\ref{eq:Fermi_distribution_two-fluid}) we
express the mass current in terms of $u$ and $\delta\mu$,
\begin{equation}
  \label{eq:mass_current}
  j=\rho u + \frac{m}{\pi\hbar}\delta\mu.
\end{equation}
The form of the first term is dictated by the Galilean invariance of
the system.  The second term is the mass current analog of the
well-known Landauer formula $I=(e^2/\pi\hbar)V$ for the electric
current $I=ej/m$ in terms of voltage $V=\delta\mu/e$.  Thus
Eq.~(\ref{eq:mass_current}) yields $j-\rho v_n=(m/\pi\hbar)\delta\mu$.

To obtain the dissipation rate in the Fermi gas, we repeat the steps
leading to Eq.~(\ref{eq:dot-n_result}) for $\dot n_p$, while using the
unperturbed distribution $n_p^{(0)}$ in the form
(\ref{eq:Fermi_distribution_two-fluid}) and allowing for small
gradients $\partial_x u$ and $\partial_x\delta\mu$.  To linear order
in the gradients we obtain
\begin{eqnarray}
  \label{eq:dot-n_result_two-fluid}
  \dot n_p&=&\frac{1}{4\mu T} n_p^{(0)}\Big(1-n_p^{(0)}\Big)\!
               \bigg[
               v_F^2(|p|-p_F)^2-\frac{\pi^2 T^2}{3}
               \bigg]
\nonumber\\
  &&\times
  \bigg(
  \gamma\,\partial_xu - \frac{1}{2p_F}\partial_x\delta\mu
  \bigg).
\end{eqnarray}
Substituting Eq.~(\ref{eq:dot-n_result_two-fluid}) along with
$\delta n_p=-\tau_2 \dot n_p$ into Eq.~(\ref{eq:W}), we obtain the rate of
dissipation in a 1D Fermi gas in the two-fluid regime.  Replacing
$u=v_n$ and $\delta\mu=(\pi\hbar/m)(j-\rho v_n)$ in the resulting
expression gives Eq.~(\ref{eq:W_two-fluid}) with
\begin{equation}
  \label{eq:zetas}
  \zeta_1=-\frac{\zeta}{\rho\gamma},
  \quad
  \zeta_2=\zeta,
  \quad
  \zeta_3=\frac{\zeta}{(\rho\gamma)^2},
\end{equation}
where $\zeta$ is given by Eq.~(\ref{eq:zeta_result}) and we have applied
the low-temperature expression $\rho=2mp_F/\pi\hbar$.  The result
$\zeta_2=\zeta$ follows immediately from the fact that in the
single-fluid regime $\delta\mu=0$.  Indeed, in this case
Eq.~(\ref{eq:mass_current}) yields $j=\rho v_n$, and
Eq.~(\ref{eq:W_two-fluid}) is identical to
Eq.~(\ref{eq:power_dissipation}).

To assess the relative importance of $\zeta_1$, $\zeta_2$, and
$\zeta_3$, we compare the quantities $\rho \zeta_1$, $\zeta_2$, and
$\rho^2 \zeta_3$, which all have the same dimension.  In the limit of
weak interactions considered here, $\gamma\ll1$, they are very
different in magnitude: $\rho^2\zeta_3 \gg \rho|\zeta_1| \gg \zeta_2$.
This result is related to our earlier observation that the
nonequilibrium response $\dot n_p$ to a small gradient $\partial_x u$
vanishes in the absence of interactions. This subtle feature of
systems of particles with quadratic spectra does not apply to the
response to the gradient $\partial_x\delta\mu$ in the two-fluid
regime, resulting in $\rho^2\zeta_3\gg \zeta_2$.  An important
application of our result (\ref{eq:zetas}) is to understanding the
attenuation of sound modes, which in the two-fluid regime is
controlled by the parameter
$\widetilde \zeta =\zeta_2-2\rho\zeta_1 +\rho^2\zeta_3$
\cite{matveev_propagation_2018}. Our result (\ref{eq:zetas})
indicates that for weakly interacting fermions the first two
contributions are negligible, and to leading order
$\widetilde \zeta =\rho^2\zeta_3$.

We have focused on the experimentally relevant and theoretically challenging case of a quadratic single-particle spectrum. If the spectrum is not quadratic, the effect of weak interactions on the spectrum need not be considered. In this case, we expect that $\zeta$ will have a form similar to Eq.~(\ref{eq:zeta_result}) without the small parameter $\gamma$. In particular, it will have the same temperature dependence as our result. Finally, the approach presented here is also applicable to the case of spinless electrons. We have verified that the results for the bulk viscosity would be consistent with those of Ref.~\cite{matveev_viscous_2017} in the regime of weak interactions.

To summarize, we studied viscous dissipation in a 1D gas of
spin-$\frac12$ fermions.  At the lowest frequencies $\omega\ll 1/\tau$,
the gas can be described by classical hydrodynamics, and its bulk
viscosity is given by our result (\ref{eq:zeta_result}).  At
frequencies above the backscattering rate,
$1/\tau\ll\omega\ll1/\tau_{\rm ex}$, two-fluid hydrodynamics is
applicable, in which the viscous effects are described by three
transport coefficients.  Our analytic expressions for these
coefficients are given by Eq.~(\ref{eq:zetas}). Our results are valid
in the broad temperature range $p_F V(0)/\hbar\ll T\ll E_F$.

\acknowledgments

The authors are grateful to A. V. Andreev and M. Pustilnik
for discussions.  Work at Argonne National Laboratory was
supported by the U.S. Department of Energy, Office of Science, Basic
Energy Sciences, Materials Sciences and Engineering Division.

\bibliography{library}

\end{document}